\documentclass[aps,reprint,nofootinbib,superscriptaddress]{revtex4}

\usepackage{graphicx}
\usepackage{amssymb}
\usepackage{amsmath}
\usepackage{bm}
\usepackage{hyperref}
\usepackage{datetime}
\usepackage[T1]{fontenc}

\newcommand{\nslash}{n\kern -0.50em /}
\newcommand{\Sslash}{\kern 0.2 em S\kern -0.50em /}

\begin{document}

\title{Signals of strong parity violation in deep inelastic scattering}

\author{Alessandro Bacchetta}
\email{alessandro.bacchetta@unipv.it \\ ORCID: 0000-0002-8824-8355}
\affiliation{Dipartimento di Fisica,  Universit\`a di Pavia, via Bassi 6,  I-27100 Pavia, Italy}
\affiliation{INFN Sezione di Pavia, via Bassi 6, I-27100 Pavia, Italy}

\author{Matteo Cerutti}
\email{matteo.cerutti@pv.infn.it \\ ORCID: 0000-0001-7238-5657}
\affiliation{Dipartimento di Fisica,  Universit\`a di Pavia, via Bassi 6,  I-27100 Pavia, Italy}
\affiliation{INFN Sezione di Pavia, via Bassi 6, I-27100 Pavia, Italy}

\author{Ludovico Manna}
\email{ludovico.manna01@ateneopv.it \\ ORCID: 0009-0003-7952-157X}
\affiliation{Dipartimento di Fisica,  Universit\`a di Pavia, via Bassi 6,  I-27100 Pavia, Italy}
\affiliation{Dipartimento di Scienza della Terra e dell'Ambiente, Universit\`a di Pavia, via Ferrata 7, I-27100 Pavia, Italy}

\author{Marco Radici}
\email{marco.radici@pv.infn.it \\ ORCID: 0000-0002-4542-9797}
\affiliation{INFN Sezione di Pavia, via Bassi 6, I-27100 Pavia, Italy}

\author{Xiaochao Zheng}
\email{xiaochao@jlab.org \\ ORCID: 0000-0001-7300-2929}
\affiliation{University of Virginia, Charlottesville, VA 22904, USA}

\begin{abstract}
We include strong parity-violating contributions to
inclusive deep inelastic scattering (DIS)
of longitudinally polarized leptons off an unpolarized target. 
At variance with standard
 results, we obtain nonvanishing parity-violating structure functions 
 in the case of pure photon exchange. The addition of these strong parity-violating contributions improves the description of existing experimental data on DIS parity-violating asymmetries. 
 We find the size of these contributions small but exhibiting a deviation from zero of about 1.5 $\sigma$. The associated $p$-value is 0.063, indicating that the probability of making an error by rejecting the hypothesis of no parity-violating contributions is 6.3\%, which is small but not negligible. 
Further improvement on the limit of the strong parity-violation can be expected from the future SoLID program at Jefferson Lab. 
%
\end{abstract}

\date{\today, \currenttime}

\maketitle
    
\section{Introduction}

Would the internal structure of the proton be identical in a mirrored world? 
According to the Standard Model (SM), it should remain the same, because Quantum ChromoDynamics (QCD) is 
invariant under parity (P) transformations.
However, there is no first principle to guarantee parity invariance in strong interactions, and
in this article we consider the impact of its violation 
(i.e., ``strong P violation'') on the internal structure of the nucleons, and 
we show that a virtual photon probing an
unpolarized proton may see more left-handed than right-handed quarks. 

The observation of strong P violation would have far-fetched consequences beyond the
specific problem of studying 
the internal structure of the nucleons.
The violation of charge-parity (CP) symmetry is the most plausible explanation of the
matter-antimatter imbalance in the Universe, but CP violation in the
SM electroweak sector alone is not sufficient to justify such an asymmetry.
An alternative solution is to go beyond the SM and introduce 
CP violation in the QCD Lagrangian.
This can be done, e.g., with the addition of
the so-called ``$\theta$-term,'' which is however constrained to be extremely
small by experimental evidence. Other higher-dimensional CP-violating operators
have been taken into consideration in the context of SM Effective Field
Theory (SMEFT) framework, which is based on the assumption that effective low-energy modifications can arise from new physics
at significantly heavier scales (see, e.g., Refs.~\cite{Brivio:2017vri,Boughezal:2021kla,USQCD:2022mmc}). 
All these extensions should give observable effects. The most studied one is the generation of permanent electric dipole moments in various particles (see, e.g., \cite{Blinov:2022tfy} and
references therein).
In the last thirty years, no experiment has been
able to observe non-zero strong CP violation. 

In this article, we show that strong P violation could be observable also in the details of the internal structure of the nucleon. 
The relaxation of P invariance
leads to the introduction of new terms in the  hadronic tensor for the Deep-Inelastic Scattering (DIS) process, generated by new 
Parton Distribution Functions (PDFs).
The new P-odd terms can be either C-even (CP-odd) or C-odd (CP-even).

We focus our interest mainly on DIS of a longitudinally polarized lepton beam off an unpolarized proton or nuclear target and we briefly discuss the case of DIS of an unpolarized lepton beam off a longitudinally polarized target.
In the first case, a new contribution in the structure function $F_3$ appears in the pure photon-exchange channel. Such contribution 
would be non-zero only with the presence of a new strong P-violating (PV) PDF. 
In order to estimate the size of such parton density, we perform a fit
to available experimental data from HERA, SLAC, and Jefferson Lab (JLab) that can be sensitive to
this kind of PV effects.

Our study could open the door for new analyses of  experimental data from DIS processes with different lepton beam polarizations and charges. 
In particular, we perform impact studies of future measurements described in the Solenoidal Large Intensity Device (SoLID) program~\cite{JeffersonLabSoLID:2022iod} at JLab 12 GeV, 
and the future Electron Ion Collider (EIC)~\cite{AbdulKhalek:2021gbh,Boughezal:2022pmb}.

\section{Formalism}

The expression of the cross section for neutral-current inclusive DIS
with an initial electron or positron off an unpolarized target reads  
\begin{equation}
\frac{d^2 \sigma}{dx_B dy} = \frac{2 \pi \alpha^2}{x_B y Q^2}\Biggl[ \Bigl(Y_+ + R^2 y^2/2 \Bigr) \bigl(F_{2,UU}+ \lambda F_{2,LU} \bigr) - y^2 \bigl(F_{L,UU} + \lambda F_{L,LU} \bigr) - {Y_-} \bigl(x_B F_{3,UU} + \lambda x_B F_{3,LU}\bigr) \Biggr],
\label{e:nc_cross}
\end{equation}
where $\alpha$ is the fine structure constant, $y$ is the inelasticity, $x_B$ is the Bjorken variable, $Q^2$ is the 
negative of the 4-momentum transfer squared of the scattering, 
$R = 2Mx_B/Q$, and 
$Y_\pm = 1\pm(1-y)^2$.
For convenience, we explicitly separated the terms proportional to $\lambda$, the helicity of the electron or positron. More details about the derivation of this expression are given in Appendix~\ref{a:xsec}. 

The structure functions involved 
in Eq.~\eqref{e:nc_cross} can be expressed as
\begin{align}
F_{2,UU}(x_B, Q^2) &= F_2^{(\gamma)} - g_V^e \eta_{\gamma Z}  F_2^{(\gamma Z)} + \bigl({g_V^e}^2+{g_A^e}^2\bigr) \eta_{Z}  F_2^{(Z)} \, , 
\label{e:F2UU}
\\
F_{2,LU}(x_B, Q^2) &=  g_A^e \eta_{\gamma Z}  F_2^{(\gamma Z)} - 2 g_V^e g_A^e \eta_{Z}  F_2^{(Z)} \, ,  
\label{e:F2LU}
\\
F_{3,UU}(x_B, Q^2) &= g_A^e \eta_{\gamma Z} F_3^{(\gamma Z)} - 2 g_V^e g_A^e \eta_{Z} F_3^{(Z)} \, , 
\label{e:F3UU} \\
F_{3,LU}^{}(x_B, Q^2) &= F_3^{(\gamma)} - g_V^e \eta_{\gamma Z} F_3^{(\gamma Z)} + \bigl({g_V^e}^2+{g_A^e}^2\bigr) \eta_{Z} F_3^{(Z)} \, ,
\label{e:F3LU}
\end{align}
where  
\begin{align}
  \eta_{\gamma Z} & =
  \biggl(\frac{G_F M_Z^2}{2 \sqrt{2} \pi \alpha} \biggr)
  \biggl(\frac{Q^2}{ Q^2 + M_Z^2} \biggr),
  &
  \eta_{Z} &= \eta_{\gamma Z}^2~,
\end{align}
with $G_F$ the Fermi constant and $M_Z$ the $Z^0$ mass.  
The $g_{V,A}^e$ are electron's neutral weak couplings to the $Z^0$. 
For positron scattering, one would change the sign of terms containing $g_A^e$, affecting Eqs.~\eqref{e:F2LU},\eqref{e:F3UU}.
The two structure functions $F_{L,UU}$ and $F_{L,LU}$ have the same decomposition as the corresponding $F_2$ ones, Eqs.~\eqref{e:F2UU} and \eqref{e:F2LU}.

The above results correspond to the standard literature (see, e.g., Ref.~\cite{Anderson:2023hhk} and the PDG
review \cite{ParticleDataGroup:2018ovx}).
The only difference is that in the SM there is no contribution to the $F_3$ structure function from pure-$\gamma$
exchange: $F_3^{(\gamma)}$ is the new ingredient in our analysis.

The observable that is affected by this additional term is the parity-violating asymmetry 
\begin{align}
  A_{\rm PV} & \equiv \frac{d\sigma (\lambda=1) - d\sigma (\lambda=-1)}{d\sigma (\lambda=1) + d\sigma (\lambda=-1)} \nonumber
  \\
  \label{e:epspinasy}
  &
= \frac{\left( Y_+ + \frac{R^2 y^2}{2} \right)\, F_{2,LU} - y^2 F_{L,LU} - {Y_-}\, x_B F_{3,LU} }{\left( Y_+ + \frac{R^2 y^2}{2} \right)\, F_{2,UU} - y^2 F_{L,UU} - {Y_-} \, x_B F_{3,UU}  }.
\end{align}
  
The structure functions can be written in terms of PDFs, which stem from the decomposition of the quark correlator. In particular, if we consider the correlator for unpolarized
nucleons and if we include strong PV terms,
the general expression at leading twist is~\footnote{A
  similar decomposition for the transverse-momentum-dependent correlator was
  studied in Ref.~\cite{Yang:2019hxu}. Our function $g_{1}^{{\rm PV}}$
  corresponds to the integral of the function $u_1$ in that reference.}
\begin{equation}
  \begin{split}
  \label{e:phi}
  \Phi^q&(x,Q^2)= \biggl\{  f_1^q(x,Q^2)
  + g_{1}^{{\rm PV} q} (x,Q^2)  \gamma_5
  \biggr\}\frac{\nslash_+}{2}
  , 
  \end{split}
\end{equation}
where $x$ is the light-cone momentum fraction carried by quarks (neglecting target mass corrections, $x \approx x_B$). 
The second term
is ignored in the SM. 
It contains the PDF $g_{1}^{{\rm PV}}$, which describes the difference in the probability to find right-handed vs. left-handed quarks inside an unpolarized proton.
It is P-odd and CP-even. Its behavior under QCD evolution is the same as the helicity distribution $g_1$.
The integral of this function 
is connected to the anapole moment of the proton and nuclei (see, e.g.,
Refs.~\cite{Flambaum:1984fc,Wood:1997zq,Zhu:2000gn,Young:2006jc}), which could contribute to PV effects in electron-proton elastic scattering experiments~\cite{SAMPLE:2000ptk,Qweak:2018tjf}. 

Neglecting strong P violation and corrections in the strong coupling constant $\alpha_S$,
but including target mass corrections (see, \textit{e.g.}, Ref.~\cite{Ruiz:2023ozv} for a recent review), the
structure function $F_3$ can be written in terms of the
PDF $f_1$~\cite{ParticleDataGroup:2018ovx}, evaluated at
the Nachtmann variable~\cite{Ruiz:2023ozv} 
\begin{equation}
    x_N = \frac{2 x_B}{1 + \sqrt{1 + R^2}} \, .
\end{equation}
For convenience, in the following we avoid
  explicitly writing the arguments of the PDFs. The detailed expression of the structure functions $F_3$ for each channel is
\begin{align}
  F_3^{(\gamma)} (x_B , Q^2) &= 0,
  \\
  \begin{split}
    F_3^{(\gamma Z)} (x_B , Q^2) &= \frac{1}{\sqrt{1+R^2}} 
    \sum_q 2 e_q g_A^q
    f_{1}^{(q-\bar{q})}, \label{e:F3gZ}
  \end{split}
  \\
    F_3^{(Z)} (x_B , Q^2) &= \frac{1}{\sqrt{1+R^2}} 
  \sum_q 2g_V^qg_A^q f_{1}^{(q-\bar{q})},
\end{align}
where $e_q$ is the quark charge,
$g_{V,A}^q$ are quark's neutral weak couplings to the $Z^0$, and $R$ has been defined below Eq.~\eqref{e:nc_cross}. 
Note that we use the simplified notation $f_1^{(q\pm\bar q)}=f_1^q\pm f_1^{\bar q}$ and similarly for the $g_1$'s below. 
Including strong P violation, 
the standard results are
modified by the following additional contributions:
\begin{align}
 \Delta F_3^{(\gamma)} (x_B , Q^2) &= - 
             \frac{1}{\sqrt{1+R^2}}
 \sum_q e_q^2
    g_{1}^{{\rm PV} (q+\bar{q})},
  \\
  \Delta  F_3^{(\gamma Z)} (x_B , Q^2) &= -
            \frac{1}{\sqrt{1+R^2}} 
    \sum_q 2 e_q
      g_V^q g_{1}^{{\rm PV} (q+\bar{q})}
        ,  \\
    \Delta F_3^{(Z)} (x_B , Q^2) &= -
            \frac{1}{\sqrt{1+R^2}} 
  \sum_q
  \bigl(g_V^{q2}+ g_A^{q2}\bigr) g_{1}^{{\rm PV} (q+\bar{q})}
        . \label{e:DF3Z}
\end{align}

The detailed SM expression of the structure function $F_2$ for each channel is
\begin{align}
 F_2^{(\gamma)} (x_B , Q^2) &= 
 \sum_q e_q^2
    f_1^{(q+\bar{q})},
  \\
  F_2^{(\gamma Z)} (x_B , Q^2) &=
    \sum_q 2 e_q
      g_V^q f_1^{(q+\bar{q})}
        ,  \\
    F_2^{(Z)} (x_B , Q^2) &=
  \sum_q
  \bigl(g_V^{q2}+ g_A^{q2}\bigr) f_1^{(q+\bar{q})}. \label{e:DF3Z}
\end{align}
Note that also the structure function $F_2$ gets modified by the 
inclusion of strong P-violation contributions:
\begin{align}
 \Delta F_2^{(\gamma)} (x_B , Q^2) &= 0,
  \\
  \begin{split}
   \Delta  F_2^{(\gamma Z)} (x_B , Q^2) &= -
    \sum_q 2 e_q
      g_A^q x_B g_{1}^{{\rm PV} (q-\bar{q})}
        ,
  \end{split}
  \\
  \begin{split}
    \Delta F_2^{(Z)} (x_B, Q^2) &= -
  \sum_q
       2 g_V^q g_A^q x_B g_{1}^{{\rm PV} (q-\bar{q})}
        .
  \end{split}
\end{align}

For completeness, we discuss the case of a polarized nucleon. The correlator becomes~\footnote{Our
  functions $f_{1L}^{{\rm PV}}$ and $e_{1T}^{\rm PV}$
  correspond to the integral of the functions $v_{1L}$ and $w_{1T}$ in
  Ref.~\cite{Yang:2019hxu}, respectively.}
\begin{equation}
  \begin{split}
  \Phi^q (x,Q^2)= \biggl\{  & f_1^q(x,Q^2)
  + g_{1}^{{\rm PV} q} (x,Q^2)  \gamma_5
  + S_L \Bigl( g_{1}^q(x,Q^2) \gamma_5
  +f_{1L}^{{\rm PV} q} (x,Q^2) \Bigr)
  \\
  & -  \Sslash_T \Bigl( h_{1}^q(x,Q^2) \gamma_5
  -e_{1T}^{{\rm PV} q} (x,Q^2) \Bigr)
  \biggr\} \frac{\nslash_+}{2} \, .
  \end{split}
\end{equation}
The fourth term, $f_{1L}^{{\rm PV}}$, is P-odd, CP-odd and
should be connected to the electric dipole moment of the proton.

In the case of DIS off longitudinally polarized protons, the $g_5$ structure
function can be
introduced~\cite{ParticleDataGroup:2018ovx}.
In this case, the inclusion of strong PV terms will generate a difference from the weak parity violation standard results, with the additional term: 
\begin{equation}
  \Delta  g_5 (x_B , Q^2)
  \approx
  \Delta  g_5^{(\gamma)} (x_B , Q^2) = \frac{1}{2}\sum_q e_q^2
 f_{1L}^{{\rm PV} (q-\bar{q})}.
\end{equation}

\section{Compatibility of experimental data with strong parity violation}

There are currently no models that generate strong P-violating PDFs.
Modifications of QCD through the inclusion of a $\theta$-term would lead to
extremely small effects, since the value of the $\theta$ parameter is
constrained by measurements of the neutron electric dipole moment. 
Other higher-dimensional P-violating SMEFT operators could possibly generate nonzero P-violating PDFs.
A model with a topologically non-trivial QCD background has been
used to generate transverse-momentum-dependent fragmentation
functions~\cite{Kang:2010qx}. 
The inclusion of electroweak
corrections in the evolution of PDFs can also produce PV PDFs.
However, given that  QED corrections are below 1\%~\cite{Ball:2013hta,Manohar:2016nzj}, PV contributions would be smaller by a factor $Q^2 / M_Z^2$, thus 
negligible at low energies.

In order to obtain a first estimate of the size of the newly introduced PV PDFs, we assume they are proportional to
their parity-even counterparts, i.e., $g_1^{\rm PV} = a\, g_1$, with $a$ being a
very small number. This leads to
\begin{align}
 \Delta F_3^{(\gamma)} (x_B , Q^2) &= - \frac{a}{\sqrt{1+R^2}} 
 \sum_q e_q^2
 g_{1}^{ (q+\bar{q})}.
\end{align}

The total
contributions to be added to the
standard expression of 
$F_3$ are
\begin{align}
    \Delta F_{3,UU}&(x_B , Q^2) = 
                - \frac{a}{\sqrt{1+R^2}} \biggl(
g_A^e \eta_{\gamma Z} \sum_q 2 e_q g_V^q g_{1}^{ (q+\bar{q})} - 2 g_V^e g_A^e \eta_{Z} \sum_q \bigl({g_V^q}^2+ {g_A^q}^2\bigr) g_{1}^{ (q+\bar{q})}
    \biggr) ,
\\
    \Delta F_{3,LU}&(x_B , Q^2) = -  
                \frac{a}{\sqrt{1+R^2}} \biggl(
    \sum_q e_q^2 g_{1}^{ (q+\bar{q})} - g_V^e \eta_{\gamma Z} \sum_q 2 e_q g_V^q g_{1}^{ (q+\bar{q})}  + \bigl({g_V^e}^2+{g_A^e}^2\bigr) \eta_{Z} \sum_q \bigl({g_V^q}^2 + {g_A^q}^2 \bigr) g_{1}^{ (q+\bar{q})}
    \biggr)
    . \nonumber
\end{align}
Moreover, the total
contributions to be added to 
$F_2$ are 
\begin{align}
\Delta F_{2,UU}&(x_B , Q^2) = g_V^e \eta_{\gamma Z} \, a\,  \sum_q 2 e_q g_A^q x_B g_{1}^{ (q-\bar{q})}  - \bigl({g_V^e}^2+{g_A^e}^2\bigr) \eta_{Z} \, a\, \sum_q 2 g_V^{q} g_A^{q} x_B g_{1}^{ (q-\bar{q})} ,
\\
\Delta F_{2,LU}&(x_B , Q^2) = - g_A^e \eta_{\gamma Z} \, a\,  \sum_q 2 e_q g_A^q x_B g_{1}^{ (q-\bar{q})}  + 2 g_V^e g_A^e \eta_{Z} \, a \, \sum_q 2 g_V^q g_A^q x_B g_{1}^{ (q-\bar{q})}. \nonumber
\end{align}
To estimate the possible size of the new PDFs, 
we proceed in the following way: we {\em assume} the validity of the
electroweak sector of the Standard Model and we attribute any discrepancy
between low-energy measurements and predictions {\em entirely} to the newly introduced
PDFs. In other words, we estimate how large
the PV PDFs should be to be compatible with low-energy
measurements.

We fit the theoretical predictions at NLO for the electron and positron PV asymmetries of Eq.~\eqref{e:epspinasy} to the DIS experimental data from HERA~\cite{H1:2018mkk} with proton beams (274 experimental points), and from JLab 6 GeV PVDIS~\cite{PVDIS:2014cmd,Wang:2014guo} (2 points) and the SLAC E122 experiment~\cite{Prescott:1979dh} (11 points) with a deuterium fixed target~\footnote{In first approximation, the deuterium target is described as a incoherent sum of free nucleons (one proton and one neutron)}. 
In Tab.\ref{t:data}, we list details of the data sets included in our analysis.
\begin{table}[h]
\footnotesize
\begin{center}
\renewcommand{\tabcolsep}{0.4pc}
\renewcommand{\arraystretch}{1.2}
\begin{tabular}{|c|c|c|c|c|c|c|c|}
  \hline
  Experiment & $N_{\rm dat}$ & Observable  & Hadron &  $\sqrt{s}$ [GeV]& $Q^2$ [GeV$^2$] &$y$ & Ref. \\
  \hline
  \hline
  HERA & 136 & $A_{\rm PV}$ for $e^+$ & proton & 319 & 120 - 30000  & 0.033 - 0.9 & \cite{H1:2018mkk} \\
  \hline
  HERA & 138 & $A_{\rm PV}$ for $e^-$ & proton & 319 & 120 - 30000  & 0.033 - 0.9 & \cite{H1:2018mkk} \\
  \hline
  JLab PVDIS & 2 & $A_{\rm PV}$ for $e^-$ & deuterium &  4.77  & 1.085; 1.901  & 0.20; 0.28 & \cite{PVDIS:2014cmd} \\
  \hline
  SLAC E122 & 11 & $A_{\rm PV}$ for $e^-$ & deuterium &  5.5 - 6.5  & 0.92 - 1.96  & 0.15 - 0.36 & \cite{Prescott:1979dh} \\
  \hline
  \hline
  Total & 287 & & & & & &\\
  \hline
\end{tabular}
\caption{
  Breakdown of the data sets considered in this analysis.
  For each data set, the table includes information on: the number of data
  points ($N_{\rm dat}$), the measured observable, the initial-state hadron, the center-of-mass energy $\sqrt{s}$, the covered range(s) in $Q^2$, the inelasticity $y$, and the published reference.
  The total number of data points amounts to 287.
  }
\label{t:data}
\end{center}
\end{table}
In all, we analyze 287 experimental data points, 136 for positron asymmetry and 151 for electron
asymmetry. 
In a future study, it could be interesting to 
estimate
the effect of the new PV PDF $g_1^{PV}$ also
in Drell--Yan processes. 

The value of the energy scale $Q^2$ is very small for E122 and JLab PVDIS data sets ($Q^2 \simeq 1-2$ GeV$^2$).
For them, 
the inclusion of target mass corrections 
(see Eq.~\eqref{e:nc_cross} and Eqs.~\eqref{e:F3gZ}-\eqref{e:DF3Z}) 
has a significant effect, 
while it does not modify the results for HERA data. 
Moreover, 
to be consistent with Refs.~\cite{PVDIS:2014cmd,Wang:2014guo}
we introduce electroweak radiative corrections according to Ref.~\cite{Erler:2013xha}.
These corrections could be included also at higher $Q^2$ (using, e.g., the Djangoh event generator~\cite{Aschenauer:2013iia}) but they are small compared to the experimental uncertainties. 

In our analysis, we choose the NNPDF4.0~\cite{NNPDF:2021njg} and the NNPDFpol1.1~\cite{Nocera:2014gqa} sets of unpolarized and polarized PDFs, respectively. We include the full set of PDF replicas to account for their uncertainty, which is interpreted as a source of systematic theoretical error to be added to the experimental systematic error. We checked that this theoretical error is much smaller than the experimental errors (of order 1\% or less). We use PDF sets at NLO accuracy and compute the structure functions $F_2$, $F_L$ and $F_3$ at $\mathcal{O}(\alpha_s)$. 

The error analysis is performed with the so-called bootstrap method,
which consists in fitting an ensemble of Monte Carlo (MC) replicas of the experimental data.
We generate
100 replicas of the experimental data 
and we relate each one of them to a single replica of unpolarized and polarized PDFs. In this way,
we obtain a distribution of 100 values for the fit parameter $a$.

The resulting quality of the fit is shown in Table~\ref{t:chitable}, where the $\chi^2$ per number of 
data points $N_{\rm dat}$ are provided for each of the considered experimental data sets,
including one standard deviation from the full ensemble of replicas.
We indicate also the values of $\chi^2 / N_{\rm dat}$ obtained with the SM predictions.

\begin{table}[h]
\footnotesize
\begin{center}
\renewcommand{\tabcolsep}{0.4pc}
\renewcommand{\arraystretch}{1.2}
\begin{tabular}{|l|c|c|c|}
  \hline
  \multicolumn{2}{|c|}{ } & \multicolumn{1}{|c|}{SM predictions} & \multicolumn{1}{|c|}{Our analysis}  \\
  \hline
  Data set & $N_{\rm dat}$ & $\chi^2 / N_{\rm dat}$  &  $\chi^2 / N_{\rm dat}$ \\
  \hline
  \hline
  HERA $e^+$ (p) & 136 & 1.12 $\pm$ 0.01 & 1.12 $\pm$ 0.01 \\
  \hline
  HERA $e^-$ (p) & 138 & 0.98 $\pm$ 0.01 & 0.98 $\pm$ 0.01 \\
  \hline
  JLab PVDIS $e^-$ (d) & 2 & 0.67 $\pm$ 0.12 & 0.42 $\pm$ 0.40 \\
  \hline
  SLAC E122 $e^-$ (d) & 11 & 0.97 $\pm$ 0.01 & 0.94 $\pm$ 0.02 \\
  \hline
  \hline
  {\bf Total} & {\bf 287} & {\bf 1.042 $\pm$ 0.001} & {\bf 1.037 $\pm$ 0.004} \\
  \hline
\end{tabular}
\caption{
  Breakdown of the values of $\chi^2$ per number of data points $N_\mathrm{dat}$ for all
  data sets considered in our analysis (hadron targets in brackets).
  The values of $\chi^2$ and uncertainties refer to the mean value and
  one standard deviation from the ensemble of replicas
  of the experimental data.
  }
\label{t:chitable}
\end{center}
\end{table}

The mean value of the global $\chi^2 / N_{\rm dat}$ is slightly smaller than the SM result, the description is thus improved by including our model for the PV PDF. 
However, it must be noted that the values of $\chi^2$ are all close to or smaller than 1 because of large experimental errors affecting many data points.
New data with higher precision would help to better assess the impact of PV contributions. 

The resulting value for the fit parameter $a$ is 
\begin{equation}
  a = \hat{a} \pm \Delta a = (-1.01 \pm 0.66) \times 10^{-4} \, ,
\end{equation}
where $\hat{a}$ and $\Delta a$ are the mean value and
one standard deviation from the ensemble of 100 values, respectively. 
Our results indicate that the null hypothesis can be rejected with a $p\text{-value} =0.063$: the probability of making an error by rejecting the hypothesis of no parity-violating contributions is 6.3\%. It is not negligible but small. In other words, current experimental data indicate that strong P violation is statistically possible and, actually, is slightly favoured with respect to the SM description. The negative value of $a$ indicates also that in the proton there could be more left-handed quarks than right-handed ones. 
We checked that this result does not depend on the choice
of the PDF by repeating the fit with different sets and obtaining no significant differences. 

\begin{figure}[h]
\centering
\includegraphics[width=0.6\textwidth]{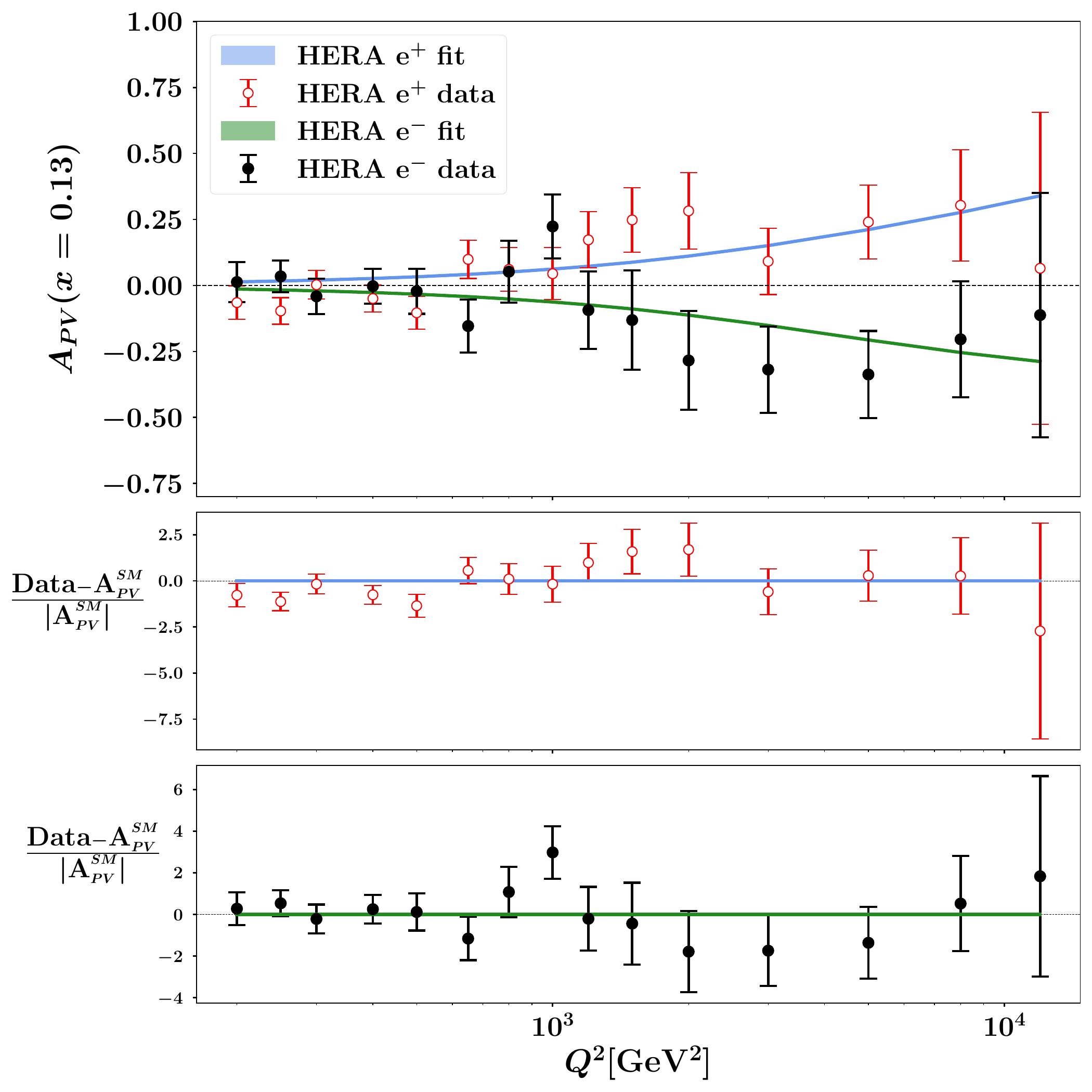}
\caption{Comparison between theoretical predictions and HERA data for $e^+p$
(open red points) and $e^-p$ (solid black points) as a function of $Q^2$ at $x=0.13$. 
Upper panel: helicity asymmetry of Eq.~\eqref{e:epspinasy};
central panel: relative difference with respect to SM predictions for $e^+$ asymmetry;
lower panel: same for $e^-$ asymmetry. 
The (barely visible) colored uncertainty bands correspond to the 68$\%$ C.L.}
\label{f:Asyplot}
\end{figure}

In Fig.~\ref{f:Asyplot}, the upper panel
shows the comparison between theoretical predictions for $A_{\rm PV}$ of Eq.~\eqref{e:epspinasy} (colored bands) and HERA data for inclusive DIS 
between a proton beam and electron $e^-$ (solid black points) or positron $e^+$ (open red points) beams, as a function of $Q^2$ at the given $x=0.13$. 
The central (lower) panel shows the relative difference of data and results of our fit with respect to SM predictions for the $e^+$ ($e^-$) asymmetry.
The (barely visible) colored uncertainty bands correspond to the 68$\%$ confidence level (C.L.), obtained by excluding the largest and smallest 16\% of the MC replicas. The narrow width of the bands reflects the small theoretical uncertainty of the PDFs used in this work. 

As can be seen from Fig.~\ref{f:Asyplot} and Tab.~\ref{t:chitable}, 
the HERA data are nicely described in our framework  although the $\chi^2$ for the $e^+$ asymmetry is slightly worse than the $e^-$ case because of the behaviour of more precise data points at the lowest $Q^2$ bins (see central panel). This may leave room for improvements of our model in a future work. 
In any case, both values of $\chi^2$ for $e^+$ and $e^-$ asymmetries are the same as in the SM framework (see Tab.~\ref{t:chitable}), indicating that PV contributions do not impact the description of HERA data: in fact, the fit is driven by the other experimental data sets with much smaller errors, described below. 

\begin{figure}[h]
\centering
\includegraphics[width=0.6\textwidth]{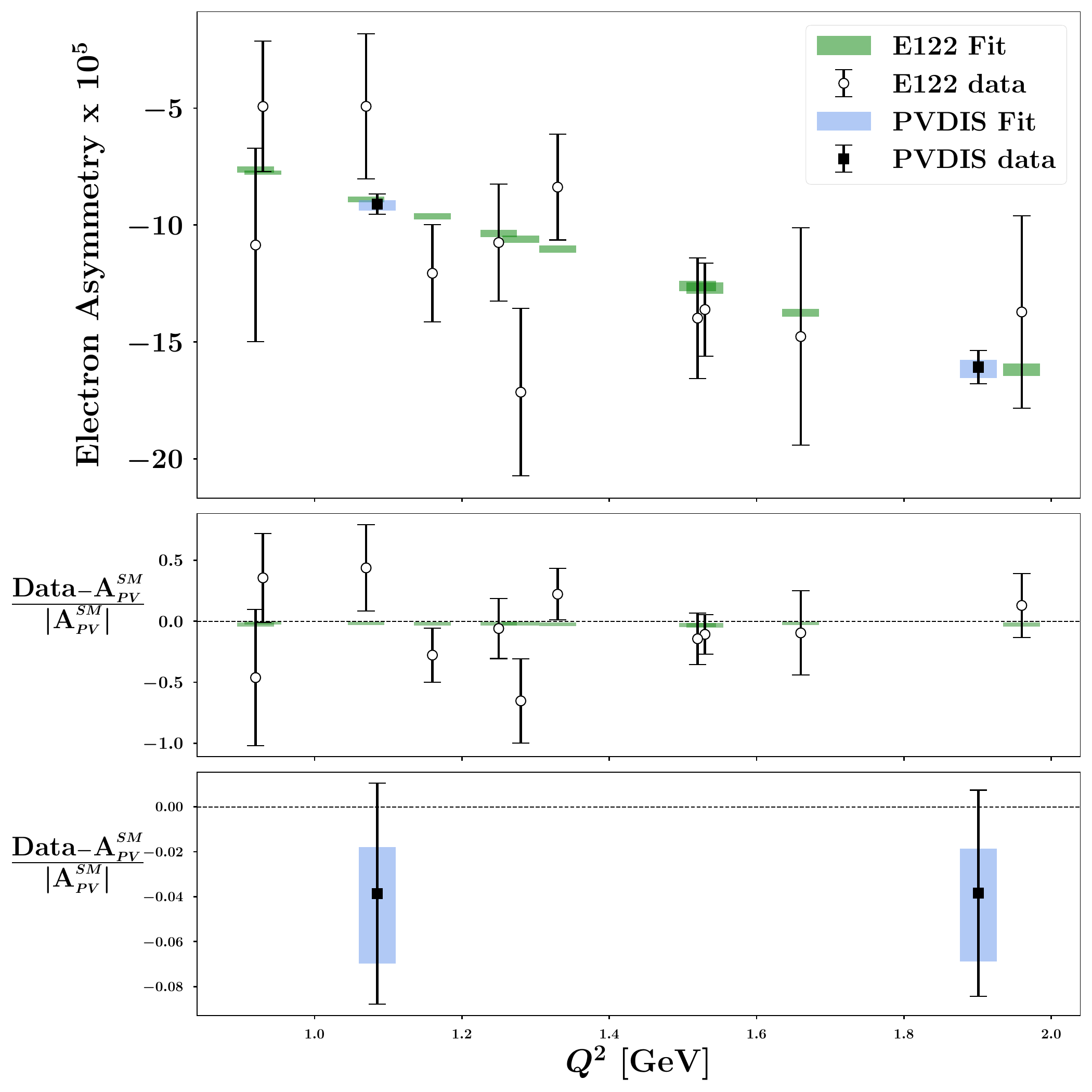}
\caption{Comparison between theoretical predictions and data from JLab 6 GeV PVDIS (solid points) and SLAC E122 (open points) experiments as a function of $Q^2$.
Upper panel: electron $e^-$ asymmetry $A_{\rm PV}$ of Eq.~\eqref{e:epspinasy}. 
Central panel: relative difference with SM predictions for the E122 asymmetry. 
Lower panel: same for the PVDIS asymmetry. Uncertainty bands correspond to the 68$\%$ C.L. }
\label{f:Asyplot_lowQ}
\end{figure}

In Fig.~\ref{f:Asyplot_lowQ}, the upper panel shows the comparison between theoretical predictions for electron $A_{\rm PV}$ of Eq.~\eqref{e:epspinasy}  (colored bands) and E122 data (open points) and JLab PVDIS data (solid points) as a function of $Q^2$. The central and the lower panel show the relative difference of data and results of our fit with respect to SM predictions for the E122 and JLab PVDIS asymmetries, respectively.
Similar to Fig.~\ref{f:Asyplot}, %
the colored uncertainty bands correspond to the 68$\%$ C.L. Both experimental data sets are nicely described in our framework. This is confirmed in Tab.~\ref{t:chitable} by the systematic improvement of the quality of the fit with respect to the SM framework. The result is particularly relevant for the JLab PVDIS data, 
due to their very small uncertainty. 

\section{Impact Study of Future Experiments}

As already remarked when commenting our results in Tab.~\ref{t:chitable}, 
further insights into the existence of PV contribution in the structure of the proton (or light nuclei) could be provided by new high-precision experimental data, such as the ones expected from the upcoming SoLID program~\cite{JeffersonLabSoLID:2022iod} at JLab 12 GeV (and possibly future upgrades of JLab~\cite{Accardi:2023chb}). 
SoLID is a versatile spectrometer designed specifically to achieve high precision measurements on a variety of topics. The nominal running conditions for SoLID's PVDIS measurement~\cite{JLabPR:PVDIS_solid} include a 50~$\mu$A, 11~GeV longitudinally polarized electron beam of 85\% polarization, incident on a 40-cm liquid deuterium or liquid hydrogen target. The pseudodata used in our study are based on PVDIS measurement with either 120~days of beam (at 100\% efficiency) on the deuterium target, or 90~days on the hydrogen target. 
The relative systematic uncertainty $\delta A_{\rm PV}{\rm (syst)}/A_{\rm PV}^{\rm SM}$ consists of an uncorrelated $0.28\%$ contribution and a $0.45\%$ contribution correlated across all kinematic bins. 

\begin{figure}[h]
\centering
\includegraphics[width=0.6\textwidth]{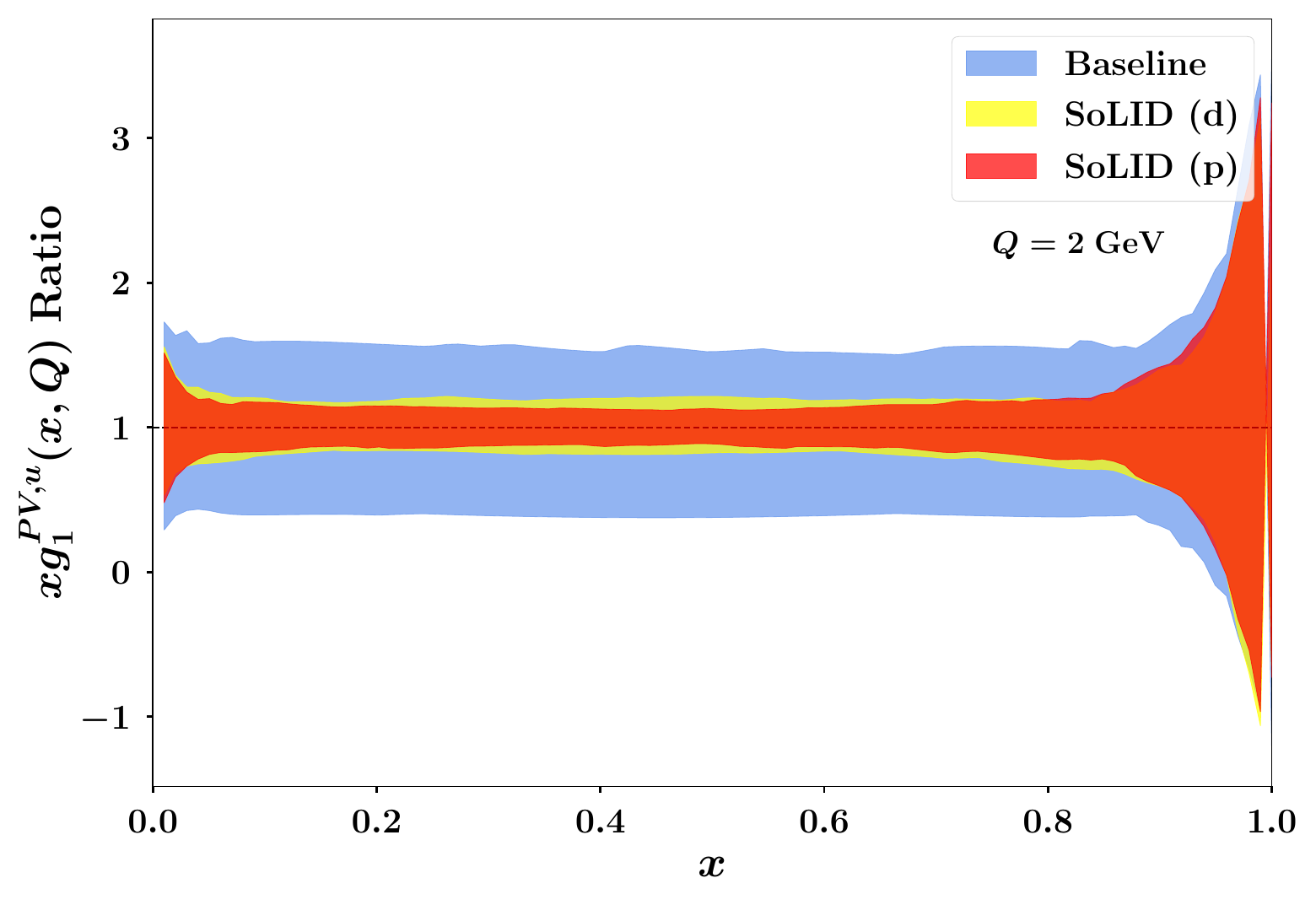}
\caption{Estimate of the impact of the SoLID pseudodata on the error bands of the $u$ quark $g_1^{\rm PV}$ in the proton in $x$ space at $Q = 2$ GeV, based on the baseline analysis presented in this paper. Purple bands: current 68\% C.L. uncertainties on $g_1^{\rm PV}$ from the baseline fit. Yellow bands: 68\% C.L. uncertainties after the inclusion of SoLID pseudodata on deuteron target. Red bands: 68\% C.L. uncertainties after the inclusion of SoLID pseudodata on proton target.}
\label{f:impact}
\end{figure}

In order to estimate the reduction of the uncertainties on the $g_1^{\rm PV}$ PDF, we generate pseudodata by calculating the theoretical predictions for the $A_{\rm PV}$ asymmetry and include the experimental uncertainties based on the running conditions given above. 
%
We consider 
the SoLID pseudodata on either deuteron or proton target and we impose a (conservative) cut 
$x_B < 0.5$, 
to avoid the kinematic region where nuclear corrections for deuteron target and higher-twist contributions become relevant.
The impact of the SoLID pseudodata on the fitted parameter $a$ is reported in Tab.~\ref{t:impact}.
\begin{table}[h]
\footnotesize
\begin{center}
\renewcommand{\tabcolsep}{0.4pc}
\renewcommand{\arraystretch}{1.2}
\begin{tabular}{|l|c|c|}
  \hline
  Fit & $a \ (10^{-4})$ & $\delta a \ (10^{-4})$  \\
  \hline
  Baseline & 1.01 & 0.66 \\
  \hline
  Baseline + SoLID (d) & 1.01 & 0.21 \\
  \hline
  Baseline + SoLID (p) & 1.01 & 0.15 \\
  \hline
\end{tabular}
\caption{
  Values of parameter $a$ and its uncertainty $\delta a$  
  related to baseline fit and the impact studies of SoLID pseudodata on deuteron (d) or proton (p) targets. 
  }
\label{t:impact}
\end{center}
\end{table}
We obtain at least a factor 3 reduction of the uncertainty with a slightly larger impact when using the proton target. If the central value of the parameter $a$ were to be confirmed by future measurements, the result in Tab.~\ref{t:impact} would represent a deviation of at least $5\sigma$ from the SM result of $a = 0$. 
Figure~\ref{f:impact} shows the impact of SoLID pseudodata on $g_1^{\rm PV}$ of the up quark normalized to its central value.
We note that the uncertainty bands decrease 
in the region $0.2<x_B<0.5$, 
which is consistent with the region covered by the SoLID pseudodata. 
The reduction outside 
this range may be 
due to the low flexibility of our assumption.
A more refined model 
will be needed when SoLID data become available. 

These results indicate that 
SoLID PVDIS data will potentially put a drastic
limit on the strong PV effects. 
Furthermore, they show that the SoLID PVDIS proton measurements will be valuable not only at high $x_B$ in providing a model-independent measurement of $d/u$ PDF ratio~\cite{JeffersonLabSoLID:2022iod}, 
but also at medium $x_B$ by offering a new way to explore physics beyond SM.


We also note that more measurements on the PVDIS asymmetries are expected from the future EIC at higher energies~\cite{AbdulKhalek:2021gbh,AbdulKhalek:2022hcn}. We considered the EIC energy configurations $10\times 275$ GeV and $10\times 137$ GeV for electron-proton and electron-deuteron collisions, respectively, each corresponds to the maximum annual integrated luminosity of 100 fb$^{-1}$~\cite{Boughezal:2022pmb,AbdulKhalek:2022hcn}. However, we found that adding these EIC pseudodata brings little impact to the limit on the strong PV effects, due to the large projected uncertainties.

\section{Conclusions}
In this paper,
we explored the impact of violating QCD parity invariance (strong P violation) on nucleon structure, focusing on inclusive DIS with longitudinally polarized leptons and unpolarized targets. 
This leads to the introduction of a new P-odd and CP-even 
PDF, denoted as $g_1^{\rm PV}$, that
describes the difference in the probability to find right-handed vs.~left-handed quarks inside an unpolarized proton. This function generates a new contribution to
the structure function $F_3$ from pure photon exchange. 

To estimate the size of $g_1^{\rm PV}$, we perform a fit to relevant experimental data from HERA, SLAC, and JLab. As a preliminary model, we assume that the PV PDFs are proportional to their parity-even counterparts and we fit the proportionality constant, $a$. 
Our analysis shows that including strong PV contributions improves the description of the data. We obtain the value $a = (-1.01 \pm 0.66) \times 10^{-4}$, which indicates that there could be more left-handed quarks than right-handed ones in an unpolarized proton.
Furthermore, we carried out impact studies for the future EIC and the SoLID PVDIS measurement of JLab's 12 GeV program. 
While in the former case we don't find any significant impact, in the latter case we find that anticipated SoLID deuteron (proton) measurement will potentially provide a factor 3 (4) reduction in the uncertainty of $a$. 
Should the central value of the parameter $a$ be confirmed by SoLID measurements, it would represent at least a $5\sigma$ deviation from the SM result of no strong PV effects. 
We emphasize that detecting strong P violation could have implications beyond nucleon structure, potentially shedding light on the matter-antimatter asymmetry in the universe. 

\begin{acknowledgments}
 We gratefully acknowledge discussion with H.~Spiesberger and P.~Souder. 
 The work of X.~Zheng is supported by the U.S. Department of Energy, Office of Science, Office of Nuclear Physics under contract number DE–SC0014434. 
\end{acknowledgments}

\begin{appendix}
\section{Derivation of the cross section}
\label{a:xsec}

For completeness, in this Appendix we provide the conventions used and the analytical derivation of the cross section in Eq.~\eqref{e:nc_cross}. 

For a fast-moving lepton with initial (final) 4-momentum $k$ ($k'=k-q$) and helicity $\lambda$, the leptonic tensor for the three channels ($\gamma$ exchange, $\gamma-Z$ interference and $Z$ exchange) can be written as
\begin{align}
    L_{\mu \nu}^{(j)} &= C^{(j)} \, L_{\mu \nu}^{(\gamma)}~, \nonumber \\
    L_{\mu \nu}^{(\gamma)} &= 2 \left[ k_\mu k_\nu' + k_\mu' k_\nu - (k \cdot k') g_{\mu \nu} - i \lambda \varepsilon_{\mu \nu \alpha \beta} k^\alpha k'^\beta \right]~,
    \label{e:L_tensor}
\end{align}
where the index $j$ runs over the channels $i=\gamma, \, \gamma Z, \, Z$, and $C^{(\gamma)} = 1, \, C^{(\gamma Z)} = - (g_V^e - \lambda g_A^e), \, C^{(Z)} = (g_V^e - \lambda g_A^e)^2$~\cite{Anderson:2023hhk}. For an anti-lepton, the same formula holds but with the sign of $g_A^e$ flipped. 

For an unpolarized hadron, the hadronic tensor is given by~\cite{ParticleDataGroup:2018ovx} 
\begin{equation}
W^{\mu \nu} = \Big( - g^{\mu \nu} + \frac{q^\mu q^\nu}{q^2} \Big) F_1 + \frac{\tilde{P}^\mu \tilde{P}^\nu}{P \cdot q} F_2 + i\frac{\varepsilon^{\mu \nu \rho \sigma}}{2(P \cdot q)}  P_\rho q_\sigma\,  F_3 \, , 
\label{e:W_tensor}
\end{equation}
where $\tilde{P}^\mu = P^\mu - q^\mu \, (P\cdot q)/q^2$. 

The cross section becomes
\begin{align}
\frac{d^2\sigma}{dx_B  \, dy} &= \frac{2\pi y \alpha^2}{Q^4} \, \sum_{j=\gamma,\,\gamma Z,\, Z} \, \eta^{(j)} \, L_{\mu \nu}^{(j)} \, W^{\mu \nu} = \frac{2\pi y \alpha^2}{Q^4} \, \sum_{j=\gamma,\,\gamma Z,\, Z} \, \eta^{(j)} \, C^{(j)} \, L_{\mu \nu}^{(\gamma)} \, W^{\mu \nu} \, , 
\label{e:cross_1}
\end{align}
where 
\begin{equation}
    \eta^\gamma = 1; \quad \quad \eta^{\gamma Z} = \left( \frac{G_F M_Z^2}{2 \sqrt{2} \pi \alpha} \right) \left( \frac{Q^2}{Q^2 + M_Z^2} \right); \quad \quad \eta^Z = \left( \eta^{\gamma Z} \right)^2~,
\end{equation}
with $\alpha$ the fine structure constant, $M_Z$ the mass of the $Z^0$ boson, and $G_F$ the Fermi coupling constant. 

By neglecting the lepton mass, the contraction of the leptonic tensor $L_{\mu \nu}^{(\gamma)}$ of Eq.~\eqref{e:L_tensor} with the hadronic tensor $W^{\mu \nu}$ of Eq.~\eqref{e:W_tensor} gives
\begin{align}
L_{\mu \nu}^{(\gamma)} \, W^{\mu \nu} &= 2 \left[ Q^2 \, F_1^{(\gamma )} + \frac{Q^2}{x_B y^2}\, \left( 1-y-\frac{1}{4}\, R^2 y^2 \right) \, F_2^{(\gamma )} - \lambda \, \frac{Q^2}{2}\,\frac{2-y}{y} \, F_3^{(\gamma )} \right] \, ,
\label{e:LW_contract}
\end{align}
where $R = 2 M x_B / Q$. By including such target mass corrections, we make the replacement $2x_B  F_1 = (1 + R^2 ) F_2 - F_L$ in the above contraction and we get
\begin{align}
L_{\mu \nu}^{(\gamma)} \, W^{\mu \nu} &= \frac{Q^2}{x_B  y^2}\, \left[ \left( Y_+ + \frac{1}{2} R^2 y^2 \right) \, F_2^{(\gamma )} - y^2 \, F_L^{(\gamma )} - \lambda \, Y_- \, x_B  F_3^{(\gamma )} \right] \, , 
\label{e:LW_contract2}
\end{align}
with $Y_\pm = 1 \pm (1-y)^2$. 

If we insert the above result into Eq.~\eqref{e:cross_1} and we use the definitions of Eqs.~\eqref{e:F2UU}-\eqref{e:F3LU}, we finally get
\begin{align}
    \frac{d^2 \sigma}{dx_B  dy} = \frac{2 \pi \alpha^2}{x_B  y Q^2}\Biggl[ &\left(Y_+ + \frac{R^2 y^2}{2} \right) \bigl(F_{2,UU}+ \lambda F_{2,LU} \bigr) - y^2 \bigl(F_{L,UU} + \lambda F_{L,LU} \bigr) \nonumber \\
       & - {Y_-} \bigl(x_B F_{3,UU} + \lambda x_B F_{3,LU}\bigr) \Biggr].
  \label{e:cross_2}
\end{align}
\end{appendix}

\bibliography{pvbiblio}

\begin{thebibliography}{31}
\expandafter\ifx\csname natexlab\endcsname\relax\def\natexlab#1{#1}\fi
\expandafter\ifx\csname bibnamefont\endcsname\relax
  \def\bibnamefont#1{#1}\fi
\expandafter\ifx\csname bibfnamefont\endcsname\relax
  \def\bibfnamefont#1{#1}\fi
\expandafter\ifx\csname citenamefont\endcsname\relax
  \def\citenamefont#1{#1}\fi
\expandafter\ifx\csname url\endcsname\relax
  \def\url#1{\texttt{#1}}\fi
\expandafter\ifx\csname urlprefix\endcsname\relax\def\urlprefix{URL }\fi
\providecommand{\bibinfo}[2]{#2}
\providecommand{\eprint}[2][]{\url{#2}}

\bibitem[{\citenamefont{Brivio and Trott}(2019)}]{Brivio:2017vri}
\bibinfo{author}{\bibfnamefont{I.}~\bibnamefont{Brivio}} \bibnamefont{and}
  \bibinfo{author}{\bibfnamefont{M.}~\bibnamefont{Trott}},
  \bibinfo{journal}{Phys. Rept.} \textbf{\bibinfo{volume}{793}},
  \bibinfo{pages}{1} (\bibinfo{year}{2019}).

\bibitem[{\citenamefont{Boughezal et~al.}(2021)\citenamefont{Boughezal,
  Petriello, and Wiegand}}]{Boughezal:2021kla}
\bibinfo{author}{\bibfnamefont{R.}~\bibnamefont{Boughezal}},
  \bibinfo{author}{\bibfnamefont{F.}~\bibnamefont{Petriello}},
  \bibnamefont{and} \bibinfo{author}{\bibfnamefont{D.}~\bibnamefont{Wiegand}},
  \bibinfo{journal}{Phys. Rev. D} \textbf{\bibinfo{volume}{104}},
  \bibinfo{pages}{016005} (\bibinfo{year}{2021}).

\bibitem[{\citenamefont{Kronfeld et~al.}(2022)}]{USQCD:2022mmc}
\bibinfo{author}{\bibfnamefont{A.~S.} \bibnamefont{Kronfeld}}
  \bibnamefont{et~al.} (\bibinfo{collaboration}{USQCD}) (\bibinfo{year}{2022}),
  {arXiv:2207.07641 [hep-lat]}.

\bibitem[{\citenamefont{Blinov et~al.}(2022)\citenamefont{Blinov, Craig, Dolan,
  de~Vries, Draper, Garcia, Lillard, and Shelton}}]{Blinov:2022tfy}
\bibinfo{author}{\bibfnamefont{N.}~\bibnamefont{Blinov}},
  \bibinfo{author}{\bibfnamefont{N.}~\bibnamefont{Craig}},
  \bibinfo{author}{\bibfnamefont{M.~J.} \bibnamefont{Dolan}},
  \bibinfo{author}{\bibfnamefont{J.}~\bibnamefont{de~Vries}},
  \bibinfo{author}{\bibfnamefont{P.}~\bibnamefont{Draper}},
  \bibinfo{author}{\bibfnamefont{I.~G.} \bibnamefont{Garcia}},
  \bibinfo{author}{\bibfnamefont{B.}~\bibnamefont{Lillard}}, \bibnamefont{and}
  \bibinfo{author}{\bibfnamefont{J.}~\bibnamefont{Shelton}}, in
  \emph{\bibinfo{booktitle}{{Snowmass 2021}}} (\bibinfo{year}{2022}),
  {arXiv:2203.07218 [hep-ph]}.

\bibitem[{\citenamefont{Arrington et~al.}(2023)}]{JeffersonLabSoLID:2022iod}
\bibinfo{author}{\bibfnamefont{J.}~\bibnamefont{Arrington}}
  \bibnamefont{et~al.} (\bibinfo{collaboration}{Jefferson Lab SoLID}),
  \bibinfo{journal}{J. Phys. G} \textbf{\bibinfo{volume}{50}},
  \bibinfo{pages}{110501} (\bibinfo{year}{2023}).

\bibitem[{\citenamefont{Abdul~Khalek
  et~al.}(2022{\natexlab{a}})}]{AbdulKhalek:2021gbh}
\bibinfo{author}{\bibfnamefont{R.}~\bibnamefont{Abdul~Khalek}}
  \bibnamefont{et~al.}, \bibinfo{journal}{Nucl. Phys. A}
  \textbf{\bibinfo{volume}{1026}}, \bibinfo{pages}{122447}
  (\bibinfo{year}{2022}{\natexlab{a}}).

\bibitem[{\citenamefont{Boughezal et~al.}(2022)\citenamefont{Boughezal, Emmert,
  Kutz, Mantry, Nycz, Petriello, \c{S}im\c{s}ek, Wiegand, and
  Zheng}}]{Boughezal:2022pmb}
\bibinfo{author}{\bibfnamefont{R.}~\bibnamefont{Boughezal}},
  \bibinfo{author}{\bibfnamefont{A.}~\bibnamefont{Emmert}},
  \bibinfo{author}{\bibfnamefont{T.}~\bibnamefont{Kutz}},
  \bibinfo{author}{\bibfnamefont{S.}~\bibnamefont{Mantry}},
  \bibinfo{author}{\bibfnamefont{M.}~\bibnamefont{Nycz}},
  \bibinfo{author}{\bibfnamefont{F.}~\bibnamefont{Petriello}},
  \bibinfo{author}{\bibfnamefont{K.}~\bibnamefont{\c{S}im\c{s}ek}},
  \bibinfo{author}{\bibfnamefont{D.}~\bibnamefont{Wiegand}}, \bibnamefont{and}
  \bibinfo{author}{\bibfnamefont{X.}~\bibnamefont{Zheng}},
  \bibinfo{journal}{Phys. Rev. D} \textbf{\bibinfo{volume}{106}},
  \bibinfo{pages}{016006} (\bibinfo{year}{2022}).

\bibitem[{\citenamefont{Anderson et~al.}(2023)\citenamefont{Anderson,
  Higinbotham, Mantry, and Zheng}}]{Anderson:2023hhk}
\bibinfo{author}{\bibfnamefont{P.}~\bibnamefont{Anderson}},
  \bibinfo{author}{\bibfnamefont{D.}~\bibnamefont{Higinbotham}},
  \bibinfo{author}{\bibfnamefont{S.}~\bibnamefont{Mantry}}, \bibnamefont{and}
  \bibinfo{author}{\bibfnamefont{X.}~\bibnamefont{Zheng}}, in
  \emph{\bibinfo{booktitle}{{30th International Workshop on Deep-Inelastic
  Scattering and Related Subjects}}} (\bibinfo{year}{2023}), {arXiv:2306.00097
  [hep-ph]}.

\bibitem[{\citenamefont{Tanabashi et~al.}(2018)}]{ParticleDataGroup:2018ovx}
\bibinfo{author}{\bibfnamefont{M.}~\bibnamefont{Tanabashi}}
  \bibnamefont{et~al.} (\bibinfo{collaboration}{Particle Data Group}),
  \bibinfo{journal}{Phys. Rev. D} \textbf{\bibinfo{volume}{98}},
  \bibinfo{pages}{030001} (\bibinfo{year}{2018}).

\bibitem[{\citenamefont{Yang}(2019)}]{Yang:2019hxu}
\bibinfo{author}{\bibfnamefont{W.}~\bibnamefont{Yang}}, \bibinfo{journal}{Int.
  J. Mod. Phys. A} \textbf{\bibinfo{volume}{34}}, \bibinfo{pages}{1950145}
  (\bibinfo{year}{2019}).

\bibitem[{\citenamefont{Flambaum et~al.}(1984)\citenamefont{Flambaum,
  Khriplovich, and Sushkov}}]{Flambaum:1984fc}
\bibinfo{author}{\bibfnamefont{V.~V.} \bibnamefont{Flambaum}},
  \bibinfo{author}{\bibfnamefont{I.~B.} \bibnamefont{Khriplovich}},
  \bibnamefont{and} \bibinfo{author}{\bibfnamefont{O.~P.}
  \bibnamefont{Sushkov}}, \bibinfo{journal}{Phys. Lett. B}
  \textbf{\bibinfo{volume}{146}}, \bibinfo{pages}{367} (\bibinfo{year}{1984}).

\bibitem[{\citenamefont{Wood et~al.}(1997)\citenamefont{Wood, Bennett, Cho,
  Masterson, Roberts, Tanner, and Wieman}}]{Wood:1997zq}
\bibinfo{author}{\bibfnamefont{C.~S.} \bibnamefont{Wood}},
  \bibinfo{author}{\bibfnamefont{S.~C.} \bibnamefont{Bennett}},
  \bibinfo{author}{\bibfnamefont{D.}~\bibnamefont{Cho}},
  \bibinfo{author}{\bibfnamefont{B.~P.} \bibnamefont{Masterson}},
  \bibinfo{author}{\bibfnamefont{J.~L.} \bibnamefont{Roberts}},
  \bibinfo{author}{\bibfnamefont{C.~E.} \bibnamefont{Tanner}},
  \bibnamefont{and} \bibinfo{author}{\bibfnamefont{C.~E.}
  \bibnamefont{Wieman}}, \bibinfo{journal}{Science}
  \textbf{\bibinfo{volume}{275}}, \bibinfo{pages}{1759} (\bibinfo{year}{1997}).

\bibitem[{\citenamefont{Zhu et~al.}(2000)\citenamefont{Zhu, Puglia, Holstein,
  and Ramsey-Musolf}}]{Zhu:2000gn}
\bibinfo{author}{\bibfnamefont{S.-L.} \bibnamefont{Zhu}},
  \bibinfo{author}{\bibfnamefont{S.~J.} \bibnamefont{Puglia}},
  \bibinfo{author}{\bibfnamefont{B.~R.} \bibnamefont{Holstein}},
  \bibnamefont{and} \bibinfo{author}{\bibfnamefont{M.~J.}
  \bibnamefont{Ramsey-Musolf}}, \bibinfo{journal}{Phys. Rev. D}
  \textbf{\bibinfo{volume}{62}}, \bibinfo{pages}{033008}
  (\bibinfo{year}{2000}).

\bibitem[{\citenamefont{Young et~al.}(2006)\citenamefont{Young, Roche, Carlini,
  and Thomas}}]{Young:2006jc}
\bibinfo{author}{\bibfnamefont{R.~D.} \bibnamefont{Young}},
  \bibinfo{author}{\bibfnamefont{J.}~\bibnamefont{Roche}},
  \bibinfo{author}{\bibfnamefont{R.~D.} \bibnamefont{Carlini}},
  \bibnamefont{and} \bibinfo{author}{\bibfnamefont{A.~W.}
  \bibnamefont{Thomas}}, \bibinfo{journal}{Phys. Rev. Lett.}
  \textbf{\bibinfo{volume}{97}}, \bibinfo{pages}{102002}
  (\bibinfo{year}{2006}).

\bibitem[{\citenamefont{Hasty et~al.}(2000)}]{SAMPLE:2000ptk}
\bibinfo{author}{\bibfnamefont{R.}~\bibnamefont{Hasty}} \bibnamefont{et~al.}
  (\bibinfo{collaboration}{SAMPLE}), \bibinfo{journal}{Science}
  \textbf{\bibinfo{volume}{290}}, \bibinfo{pages}{2117} (\bibinfo{year}{2000}).

\bibitem[{\citenamefont{Androi\'c et~al.}(2018)}]{Qweak:2018tjf}
\bibinfo{author}{\bibfnamefont{D.}~\bibnamefont{Androi\'c}}
  \bibnamefont{et~al.} (\bibinfo{collaboration}{Qweak}),
  \bibinfo{journal}{Nature} \textbf{\bibinfo{volume}{557}},
  \bibinfo{pages}{207} (\bibinfo{year}{2018}).

\bibitem[{\citenamefont{Ruiz et~al.}(2023)}]{Ruiz:2023ozv}
\bibinfo{author}{\bibfnamefont{R.}~\bibnamefont{Ruiz}} \bibnamefont{et~al.}
  (\bibinfo{year}{2023}), {arXiv:2301.07715 [hep-ph]}.

\bibitem[{\citenamefont{Kang and Kharzeev}(2011)}]{Kang:2010qx}
\bibinfo{author}{\bibfnamefont{Z.-B.} \bibnamefont{Kang}} \bibnamefont{and}
  \bibinfo{author}{\bibfnamefont{D.~E.} \bibnamefont{Kharzeev}},
  \bibinfo{journal}{Phys. Rev. Lett.} \textbf{\bibinfo{volume}{106}},
  \bibinfo{pages}{042001} (\bibinfo{year}{2011}).

\bibitem[{\citenamefont{Ball et~al.}(2013)\citenamefont{Ball, Bertone,
  Carrazza, Del~Debbio, Forte, Guffanti, Hartland, and Rojo}}]{Ball:2013hta}
\bibinfo{author}{\bibfnamefont{R.~D.} \bibnamefont{Ball}},
  \bibinfo{author}{\bibfnamefont{V.}~\bibnamefont{Bertone}},
  \bibinfo{author}{\bibfnamefont{S.}~\bibnamefont{Carrazza}},
  \bibinfo{author}{\bibfnamefont{L.}~\bibnamefont{Del~Debbio}},
  \bibinfo{author}{\bibfnamefont{S.}~\bibnamefont{Forte}},
  \bibinfo{author}{\bibfnamefont{A.}~\bibnamefont{Guffanti}},
  \bibinfo{author}{\bibfnamefont{N.~P.} \bibnamefont{Hartland}},
  \bibnamefont{and} \bibinfo{author}{\bibfnamefont{J.}~\bibnamefont{Rojo}}
  (\bibinfo{collaboration}{NNPDF}), \bibinfo{journal}{Nucl. Phys. B}
  \textbf{\bibinfo{volume}{877}}, \bibinfo{pages}{290} (\bibinfo{year}{2013}).

\bibitem[{\citenamefont{Manohar et~al.}(2016)\citenamefont{Manohar, Nason,
  Salam, and Zanderighi}}]{Manohar:2016nzj}
\bibinfo{author}{\bibfnamefont{A.}~\bibnamefont{Manohar}},
  \bibinfo{author}{\bibfnamefont{P.}~\bibnamefont{Nason}},
  \bibinfo{author}{\bibfnamefont{G.~P.} \bibnamefont{Salam}}, \bibnamefont{and}
  \bibinfo{author}{\bibfnamefont{G.}~\bibnamefont{Zanderighi}},
  \bibinfo{journal}{Phys. Rev. Lett.} \textbf{\bibinfo{volume}{117}},
  \bibinfo{pages}{242002} (\bibinfo{year}{2016}).

\bibitem[{\citenamefont{Andreev et~al.}(2018)}]{H1:2018mkk}
\bibinfo{author}{\bibfnamefont{V.}~\bibnamefont{Andreev}} \bibnamefont{et~al.}
  (\bibinfo{collaboration}{H1}), \bibinfo{journal}{Eur. Phys. J. C}
  \textbf{\bibinfo{volume}{78}}, \bibinfo{pages}{777} (\bibinfo{year}{2018}).

\bibitem[{\citenamefont{Wang et~al.}(2014)}]{PVDIS:2014cmd}
\bibinfo{author}{\bibfnamefont{D.}~\bibnamefont{Wang}} \bibnamefont{et~al.}
  (\bibinfo{collaboration}{PVDIS}), \bibinfo{journal}{Nature}
  \textbf{\bibinfo{volume}{506}}, \bibinfo{pages}{67} (\bibinfo{year}{2014}).

\bibitem[{\citenamefont{Wang et~al.}(2015)}]{Wang:2014guo}
\bibinfo{author}{\bibfnamefont{D.}~\bibnamefont{Wang}} \bibnamefont{et~al.},
  \bibinfo{journal}{Phys. Rev. C} \textbf{\bibinfo{volume}{91}},
  \bibinfo{pages}{045506} (\bibinfo{year}{2015}).

\bibitem[{\citenamefont{Prescott et~al.}(1979)}]{Prescott:1979dh}
\bibinfo{author}{\bibfnamefont{C.~Y.} \bibnamefont{Prescott}}
  \bibnamefont{et~al.}, \bibinfo{journal}{Phys. Lett. B}
  \textbf{\bibinfo{volume}{84}}, \bibinfo{pages}{524} (\bibinfo{year}{1979}).

\bibitem[{\citenamefont{Erler and Su}(2013)}]{Erler:2013xha}
\bibinfo{author}{\bibfnamefont{J.}~\bibnamefont{Erler}} \bibnamefont{and}
  \bibinfo{author}{\bibfnamefont{S.}~\bibnamefont{Su}}, \bibinfo{journal}{Prog.
  Part. Nucl. Phys.} \textbf{\bibinfo{volume}{71}}, \bibinfo{pages}{119}
  (\bibinfo{year}{2013}).

\bibitem[{\citenamefont{Aschenauer et~al.}(2013)\citenamefont{Aschenauer,
  Burton, Martini, Spiesberger, and Stratmann}}]{Aschenauer:2013iia}
\bibinfo{author}{\bibfnamefont{E.~C.} \bibnamefont{Aschenauer}},
  \bibinfo{author}{\bibfnamefont{T.}~\bibnamefont{Burton}},
  \bibinfo{author}{\bibfnamefont{T.}~\bibnamefont{Martini}},
  \bibinfo{author}{\bibfnamefont{H.}~\bibnamefont{Spiesberger}},
  \bibnamefont{and}
  \bibinfo{author}{\bibfnamefont{M.}~\bibnamefont{Stratmann}},
  \bibinfo{journal}{Phys. Rev. D} \textbf{\bibinfo{volume}{88}},
  \bibinfo{pages}{114025} (\bibinfo{year}{2013}).

\bibitem[{\citenamefont{Ball et~al.}(2022)}]{NNPDF:2021njg}
\bibinfo{author}{\bibfnamefont{R.~D.} \bibnamefont{Ball}} \bibnamefont{et~al.}
  (\bibinfo{collaboration}{NNPDF}), \bibinfo{journal}{Eur. Phys. J. C}
  \textbf{\bibinfo{volume}{82}}, \bibinfo{pages}{428} (\bibinfo{year}{2022}).

\bibitem[{\citenamefont{Nocera et~al.}(2014)\citenamefont{Nocera, Ball, Forte,
  Ridolfi, and Rojo}}]{Nocera:2014gqa}
\bibinfo{author}{\bibfnamefont{E.~R.} \bibnamefont{Nocera}},
  \bibinfo{author}{\bibfnamefont{R.~D.} \bibnamefont{Ball}},
  \bibinfo{author}{\bibfnamefont{S.}~\bibnamefont{Forte}},
  \bibinfo{author}{\bibfnamefont{G.}~\bibnamefont{Ridolfi}}, \bibnamefont{and}
  \bibinfo{author}{\bibfnamefont{J.}~\bibnamefont{Rojo}}
  (\bibinfo{collaboration}{NNPDF}), \bibinfo{journal}{Nucl. Phys. B}
  \textbf{\bibinfo{volume}{887}}, \bibinfo{pages}{276} (\bibinfo{year}{2014}).

\bibitem[{\citenamefont{Accardi et~al.}(2023)}]{Accardi:2023chb}
\bibinfo{author}{\bibfnamefont{A.}~\bibnamefont{Accardi}} \bibnamefont{et~al.}
  (\bibinfo{year}{2023}), {arXiv:2306.09360 [nucl-ex]}.

\bibitem[{\citenamefont{Souder~(contact) et~al.}(2010 with 2022
  update)\citenamefont{Souder~(contact), Reimer, Zheng
  et~al.}}]{JLabPR:PVDIS_solid}
\bibinfo{author}{\bibfnamefont{P.~A.} \bibnamefont{Souder~(contact)}},
  \bibinfo{author}{\bibfnamefont{P.~E.} \bibnamefont{Reimer}},
  \bibinfo{author}{\bibfnamefont{X.}~\bibnamefont{Zheng}},
  \bibnamefont{et~al.}, \emph{\bibinfo{title}{{Precision Measurement of
  Parity-violation in Deep Inelastic Scattering Over a Broad Kinematic
  Range}}}, \bibinfo{howpublished}{Jefferson Lab Experiment E12-10-007}
  (\bibinfo{year}{2010 with 2022 update}).

\bibitem[{\citenamefont{Abdul~Khalek
  et~al.}(2022{\natexlab{b}})}]{AbdulKhalek:2022hcn}
\bibinfo{author}{\bibfnamefont{R.}~\bibnamefont{Abdul~Khalek}}
  \bibnamefont{et~al.} (\bibinfo{year}{2022}{\natexlab{b}}), {arXiv:2203.13199
  [hep-ph]}.

\end{thebibliography}
\bibliographystyle{myrevtex}

\end{document}